\documentclass[aoas,citesort,seceqn,dvips]{arximspdf}
\usepackage[vtexbibl]{}
\usepackage{graphicx}

\doi{10.1214/09-AOAS260}
\volume{3}
\issue{4}
\pubyear{2009}
\firstpage{1710}
\lastpage{1737}

\makeatletter
\newtheorem{teorem}{Theorem}[section]
\newproclaim{algorithm}{Algorithm}[section]
\newproclaim{procedure}{Procedure}[section]
\makeatother

\begin{document}
\begin{frontmatter}

\title{Improving the precision of classification trees\protect\thanksref{T1}}
\runtitle{Classification trees}
\thankstext{T1}{This material is based upon work partially supported
by U.S. Army Research Office Grants W911NF-05-1-0047 and W911NF-09-1-0205.}

\begin{aug}
\author{\fnms{Wei-Yin} \snm{Loh}\ead[label=e1]{loh@stat.wisc.edu}\corref{}}
\runauthor{W.-Y. Loh}
\affiliation{University of Wisconsin}
\address{Department of Statistics\\University of Wisconsin\\
1300 University Avenue\\Madison, Wisconsin 53706\\USA\\
\printead{e1}} 
\end{aug}

\received{\smonth{8} \syear{2008}}
\revised{\smonth{5} \syear{2009}}

%
\begin{abstract}
Besides serving as prediction models, classification trees are
useful for finding important predictor variables and identifying
interesting subgroups in the data. These functions can be
compromised by weak split selection algorithms that have variable
selection biases or that fail to search beyond local main effects at
each node of the tree. The~resulting models may include many
irrelevant variables or select too few of the important ones.
Either eventuality can lead to erroneous conclusions. Four
techniques to improve the precision of the models are proposed and
their effectiveness compared with that of other algorithms,
including tree ensembles, on real and simulated data sets.
\end{abstract}

%
\begin{keyword}
\kwd{Bagging}
\kwd{kernel density}
\kwd{discrimination}
\kwd{nearest neighbor}
\kwd{prediction}
\kwd{random forest}
\kwd{selection bias}
\kwd{variable selection}.
\end{keyword}

\end{frontmatter}

\section{Introduction} \label{intro}

Since the appearance of the AID and THAID algorithms
\cite{aid,thaid}, classification trees have offered a unique way to
model and visualize multi-dimensional data.  As such, they are more
intuitive to interpret than models that can be described only with
mathematical equations.  Interpretability, however, ensures neither
predictive accuracy nor model parsimony.  Predictive accuracy is the
probability that a model correctly classifies an independent
observation not used for model construction.  Parsimony is always
desirable in modeling---see, for example, McCullagh and Nelder
\cite{glmbook}, page~7---but it takes on increased importance here.  A
tree that involves irrelevant variables is not only more cumbersome to
interpret but also potentially misleading.

Many of the early classification tree algorithms, including THAID,
CART \cite{cart} and C4.5 \cite{c45}, search exhaustively for a
split of a node by minimizing a measure of node heterogeneity.  As a
result, if all other things are equal, variables that take more values
have a greater chance to be chosen. This selection bias can produce
overly large or overly small tree structures that obscure the
importance of the variables. Doyle \cite{doyle73} seems to be the first to
raise this issue, but solutions have begun to appear only in the last
decade or so. The~QUEST \cite{quest} and CRUISE \cite{cruise}
algorithms avoid the bias by first using \textit{F} and chi-squared
tests at each node to select the variable to split on.  CTree
\cite{party} uses permutation tests. Other approaches are proposed
in \cite{NSP04,LS06} and \cite{SBA07}.

Unbiasedness alone, however, guarantees neither predictive accuracy
nor variable selection efficiency.  To see this, consider some data on
the mammography experience (\texttt{ME}) of 412 women from Hosmer and
Lemeshow \cite{HL00}, \mbox{pages~264--287}.  The~class variable is
\texttt{ME}, with 104 women having had a mammogram within the last
year (\texttt{ME}${}={}$1), 74 having had one more than a year ago
(\texttt{ME}${}={}$2), and 234~women not having had any (\texttt{ME}${}={}$3).
[The~\texttt{ME} codes here are different from the source; they are
chosen to reflect a natural ordering.]  Table~\ref{tab:mammo:vars}
lists the five predictor variables and the values they take.  Hosmer
and Lemeshow fitted a polytomous logistic regression model for
predicting \texttt{ME} that includes all five predictor variables, but
with \texttt{SYMP} and \texttt{DETC} in the form of indicator
variables $I(\mathtt{SYMP} \leq 2)$ and $I(\mathtt{DETC} = 3)$.

\begin{table}
  \caption{Variables in mammography data}
  \label{tab:mammo:vars}
   \begin{tabular*}{\textwidth}{@{\extracolsep{\fill}}lll@{}}
   \hline
    \textbf{Name} & \textbf{Description} & \textbf{Values} \\
    \hline
    \texttt{ME} & Mammography experience & 1 (within 1 year), 2 (more \\
     &&  than 1 year), 3 (never)\\
    \texttt{SYMP} & You do not need a mammogram unless you &  1 (strongly agree), 2  \\
     &  develop symptoms &   (agree), 3 (disagree), 4  \\
      &   &    (strongly disagree) \\
    \texttt{PB} & Perceived benefit of mammography & 5--20 (lower values indicate  \\
     & &  greater benefit) \\
    \texttt{HIST} & Mother or sister with history of breast cancer & 0 (no), 1 (yes) \\
    \texttt{BSE} & Has anyone taught you how to examine your  & 0 (no), 1 (yes) \\
     &  own breasts? &  \\
    \texttt{DETC} & How likely is it that a mammogram could find  & 1 (not likely), 2 (somewhat \\
 &  a new case of breast  cancer? & likely), 3 (very likely) \\
    \hline
  \end{tabular*}
\end{table}

\begin{figure}

\includegraphics{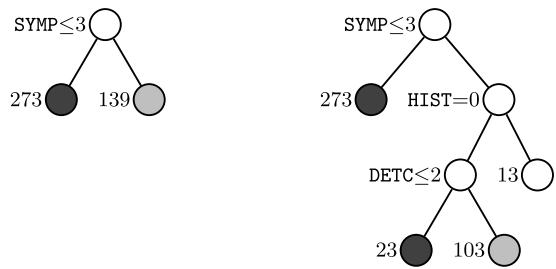}

  \caption{QUEST (left) and CRUISE (right) classification trees for
    mammography data. At each intermediate node, an observation goes
    to the left branch if and only if the condition shown there is
    satisfied.  Leaf nodes classified as class~1, 2 and 3 are colored
    white, light gray, and dark gray, respectively.  The~number on the
    left of each leaf node is the sample size.}
  \label{fig:mammo:CQ}
\end{figure}

Let $C(i|j)$ denote the cost of misclassifying as class $i$ an
observation whose true class is $j$.  Because the \texttt{ME} values
are ordered, we set $C(i|j) = |i-j|$ for $i, j = 1, 2, 3$.
Figure~\ref{fig:mammo:CQ} shows the QUEST and CRUISE models.  Each
leaf node is colored white, light gray or dark gray as the predicted
value of \texttt{ME} is 1, 2 or 3, respectively. The~QUEST tree is
too short; it splits only once and does not predict class~1.  Note
that class~1 constitutes less than 18\% of the sample and that the
Hosmer--Lemeshow model \cite{HL00}, page~277, Table~8.10, does not predict
class~1 too.  Another reason that the QUEST tree is shorter than the
CRUISE tree is because the latter tests for pairwise interactions at
each node, whereas the former does not. Thus, CRUISE can uncover more
structure than QUEST.

\begin{figure}[b]

\includegraphics{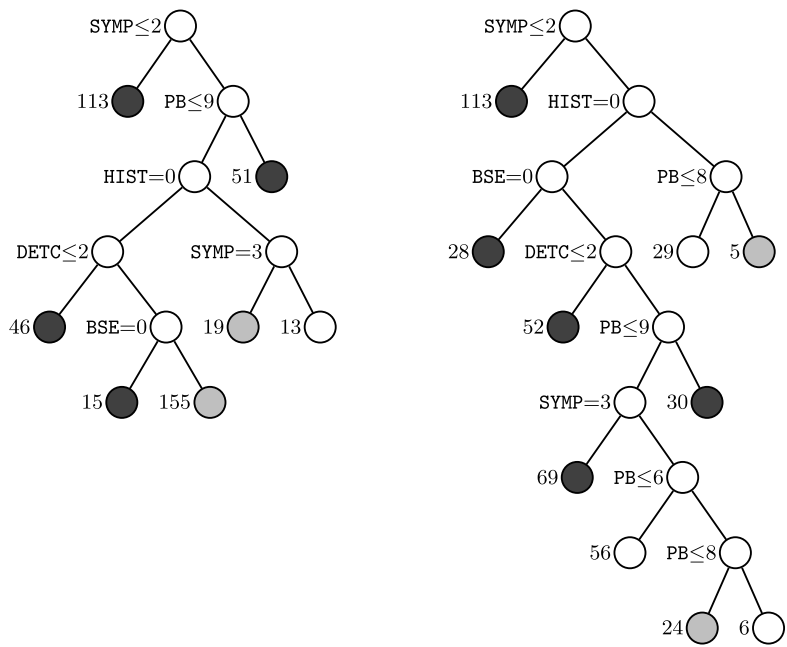}

  \caption{RPART (left) and J48 (right) classification trees for
    mammography data. At each intermediate node, an observation goes
    to the left branch if and only if the condition shown there is
    satisfied.  Leaf nodes classified as class~1, 2 and 3 are colored
    white, light gray and dark gray, respectively.  The~number on the
    left of each leaf node is the sample size.}
  \label{fig:mammo:RJ}
\end{figure}

Figure~\ref{fig:mammo:RJ} shows the corresponding RPART \cite{rpart}
and J48 \cite{weka} trees.  RPART is an implementation of CART in R
and J48 is an implementation of C4.5 in JAVA.  The~two trees have much
more structure, but the J48 tree reminds us that the comprehensibility
of a tree structure diminishes with increase in its size.  Further,
there is a hint of over-fitting in the repeated and nonmonotonic
splits on \texttt{PB} at the bottom of the J48 tree.  It is difficult
to tell which tree model has higher or lower predictive accuracy than
the Hosmer and Lemeshow model.  Empirical studies have shown that the
predictive accuracy of logistic regression is often as good as that of
classification trees for small sample sizes \cite{LLS00} but that C4.5
is more accurate as the sample size increases \cite{PPS03}.  (Note:
all the trees discussed in this article are pruned according to their
respective algorithms. Trees constructed by the new methods to be
described are pruned using CART's cost-complexity method with ten-fold
cross-validation.)

\begin{figure}[b]

\includegraphics{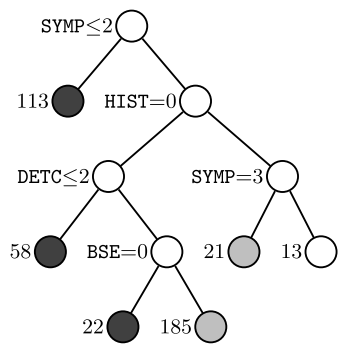}

  \caption{Tree for mammography data using the S method.  At each
    intermediate node, an observation goes to the left branch if and
    only if the condition shown there is satisfied.  Leaf nodes
    classified as class~1, 2 and 3 are colored white, light gray and
    dark gray, respectively.  The~number on the left of each leaf node
    is the sample size.}
  \label{fig:mammo:guide}
\end{figure}

Our goal here is to introduce and examine four techniques for
constructing tree models that are parsimonious and have high
predictive accuracy. We do this by controlling the search for local
interactions, employing more effective variable and split selection
strategies, and fitting nontrivial models in the nodes. The~first
technique limits the frequency of interaction tests, performing them
only if no main effect test is significant. Besides saving computation
time, this reduces the chance of selecting variables of lesser
importance.  The~second technique utilizes a two-level split search
when a significant interaction is found.  This allows greater
advantage to be taken of the information gained from interaction
tests.  The~third technique considers linear splits on pairs of
variables, if no main effect or interaction test is significant at a
node.  This can sometimes produce large gains in predictive accuracy
as well as reduction in tree size.  The~fourth technique fits a
nearest-neighbor or a kernel discriminant model, using one or two
variables, at each node.  This is useful in applications where neither
univariate nor linear splits are effective.  The~result of applying
the first three techniques to the mammography data is given in
Figure~\ref{fig:mammo:guide}. Its model complexity is in between that
of CRUISE and QUEST on one hand, and that of RPART and C4.5 on the
other.

The~rest of this article is organized as follows.
Section~\ref{sec:control-interact} discusses why and how we control
the search for interactions and illustrates the solution with an
artificial example. Section~\ref{sec:linsplits} introduces linear
splits on pairs of variables and motivates the solution with a real
example. Section~\ref{sec:bias} presents simulation results to show
that the selection bias of our method is well controlled.
Section~\ref{sec:knn} describes the use of kernel and nearest-neighbor
models on pairs of variables to fit the data in each node and
demonstrates their effectiveness with yet another example.
Section~\ref{sec:comp} compares the predictive accuracy, tree size
and execution time of the algorithms on forty-six data
sets. Section~\ref{sec:ensembles} examines the effect of using tree
ensembles and compares the results with Random Forest
\cite{randomforest01}. Section~\ref{sec:conc} concludes the
discussion.

\section{Controlled search for local interactions}
\label{sec:control-interact}
The~brevity of the QUEST tree in Figure~\ref{fig:mammo:CQ} is
attributable to its split selection strategy.  At each node, QUEST
evaluates the within-node (i.e., local) main effect of each predictor
variable by computing an ANOVA \mbox{$p$-value} for each noncategorical
variable and a Pearson chi-squared \mbox{$p$-value} for each categorical
variable. Then it selects the variable with the smallest \mbox{$p$-value} to
split the node. Although successful in avoiding selection bias, QUEST
is insensitive to (local) interaction effects within the node. If the
latter are strong and local main effects weak, the algorithm can
select the wrong variable.  The~weakness does not necessarily lead to
reduced predictive accuracy, however, because good splits may be found
farther down the tree---this explains why algorithms with selection
biases can still yield models with good predictive accuracy.

CRUISE searches over a larger number of splits at each node by
including tests of interactions between all pairs of predictors.  If
there are $K$ predictor variables, CRUISE computes $K$ main effect
$p$-values and $K(K-1)/2$ interaction $p$-values and splits the node
on the variable associated with the smallest \mbox{$p$-value}. Because there
are usually more interaction tests than main effect tests, the
smallest \mbox{$p$-value} often comes from the former. As a result, CRUISE
may select a variable with a weak main effect even though there are
other variables with stronger ones. Further, if the most significant
\mbox{$p$-value} is from an interaction between two variables, CRUISE chooses
the variable with the smaller main effect \mbox{$p$-value} and then searches
for a split on that variable alone.

\begin{figure}

\includegraphics{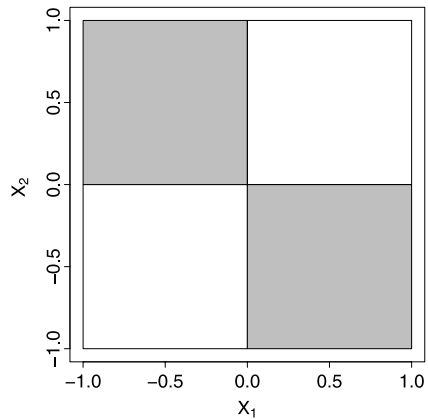}

  \caption{Class 1 observations uniformly distributed in the white
    squares and class 2 observations uniformly distributed in the
    gray squares.}
  \label{fig:rect}
\end{figure}

To see that this can create difficulties, consider an extreme example
where there are two classes and two predictor variables, $X_1$ and
$X_2$, distributed on a square as in Figure~\ref{fig:rect}. The~square
is divided equally into four sub-squares with one class located in the
two sub-squares on one diagonal and the other class in the other two
sub-squares.  The~optimal classification rule first splits the space
into two equal halves at the origin using either $X_1$ or $X_2$ and
then splits each half into two at the origin of the other variable.
Since this requires a two-level search, CRUISE will likely require
many more splits to accurately classify the data.

The~above problems can be solved by making two adjustments. First, to
prevent the interaction tests from overwhelming the main effect tests,
we carry out the former only when no main effect test achieves a
specified level of significance. Second, we perform a two-level search
for the split points whenever a significant interaction is
found. These steps are given in detail below, after the introduction
of some necessary notation.

Let $J$ denote the number of classes in the training sample and $J_t$
the number of classes in node $t$. Let $N_j$ denote the number of
class $j$ cases in the training sample and $N = \sum_j N_j$. Let
$N_j(t)$ and $N(t)$ denote the corresponding sample sizes in $t$, and
let $\pi(j)$ be the prior probability for class~$j$. The~value of
$\pi(j)$ may be specified by the user or it can be estimated from the
data, in which case it is $N_j/N$. The~estimated probability that a
class $j$ observation will land in node $t$ is $p(j,t) = \pi(j)
N_j(t)/N_j$. Define $p(t) = \sum_j p(j,t)$ and $p(j|t) = p(j,t)/p(t)$.
The~Gini impurity at $t$ is defined as $g(t) = \sum_{i \neq j}
p(i|t)p(j|t)$.

\subsection{Variable selection}
\label{sec:var-select}

Following CRUISE, we use Pearson chi-squared tests to assess the main
effects of the predictor variables. For a noncategorical predictor
variable, we discretize its values into three or four intervals,
depending on the sample size and the number of classes in the node.
But we do not convert the chi-squared values to $p$-values as CRUISE
does. Instead, for degrees of freedom (d.f.) greater than one, we use
the Wilson--Hilferty \cite{wh31} approximation to transform
each chi-squared value to a standard normal deviate and then use its
inverse transformation to convert it to a chi-squared value with one
d.f.  This technique, which is borrowed from GUIDE \cite{guide},
avoids the difficulties of computing very small $p$-values. The~detailed procedure is as follows.

\begin{procedure} \label{proc:main}
    Main effect chi-squared statistic for $X$ at node $t$:
  \begin{enumerate}[2]
  \item If $X$ is a categorical variable:
    \begin{enumerate}[(b)]
    \item[(a)] Form a contingency table with the class labels as rows and
      the categories of $X$ as columns.
    \item[(b)] Let $\nu$ be the degrees of freedom of the table after
      deleting any rows and columns with no observations. Compute the
      chi-squared statistic $\chi^2_{\nu}$ for testing
      independence. If $\nu > 1$, use the Wilson--Hilferty
      approximation twice to convert $\chi^2_{\nu}$ to the 1-d.f.
      chi-squared
      \begin{equation}
        \label{eq:wh}
        W_M(X) = \max \biggl(0, \biggl[ \frac{7}{9} + \sqrt{\nu} \biggl\{
              \biggl( \frac{\chi^2_{\nu}}{\nu} \biggr)^{1/3} - 1 +
              \frac{2}{9\nu} \biggr\} \biggr]^3 \biggr).
      \end{equation}
    \end{enumerate}

  \item If $X$ is a noncategorical variable:
    \begin{enumerate}[(b)]
    \item[(a)] Compute $\bar{x}$ and $s$, the mean and standard deviation,
      respectively, of the values of $X$ in $t$.
    \item[(b)] If $N(t) \geq 20J_t$, divide the range of $X$ into four
      intervals with boundary values $\bar{x}$ and $\bar{x} \pm
      s\sqrt{3}/2$.  Otherwise, if $N(t) < 20J_t$, divide the range of
      $X$ into three intervals with boundary values $\bar{x} \pm
      s\sqrt{3}/3$.  If $X$ has a uniform distribution, these
      boundaries will yield intervals with roughly equal numbers of
      observations.
    \item[(c)] Form a contingency table with the class labels as rows and
      the intervals as columns.
    \item[(d)] Follow step~1(b) to obtain $W_M(X)$.
    \end{enumerate}
  \end{enumerate}
\end{procedure}

Discretization in step~2(b) is needed to permit application of the
chi-squared test.  Although the boundaries are chosen for reasons of
computational expediency, our empirical experience indicates that the
particular choice is not critical for its purpose, namely,
unbiasedness in variable selection. Note that the boundaries are not
used as split points; a search for the latter is carried out
separately in Algorithm~\ref{alg:simple} below.

We apply the same idea to assess the local interaction effects of each
pair of variables, using Cartesian products of sets of values for the
columns of the chi-squared table.

\begin{procedure} \label{proc:interact}
     Interaction chi-squared statistic for a pair of variables $X_1$ and
    $X_2$ at node $t$:
  \begin{enumerate}[3.]
  \item If $X_i$ ($i = 1, 2$) is noncategorical, split its range into
    two intervals ($A_{i1}, A_{i2}$) at the sample mean $\bar{x}$ if
    $N(t) < 45J_t$, or three intervals ($A_{i1}, A_{i2}, A_{i3}$) at
    the points $\bar{x} \pm s\sqrt{3}/3$, if $N(t) \geq 45J_t$.  If
    $X_i$ ($i = 1, 2$) is categorical, let $A_{ik}$ denote the
    singleton set containing its $k$th value.
  \item Divide the $(X_1, X_2)$-space into sets $B_{k,m} = \{(x_1,
    x_2) \dvtx  x_1 \in A_{1k}, x_2 \in A_{2m} \}$, for $k, m = 1,2,
    \ldots.$
  \item Form a contingency table with the class labels as rows and
    $\{B_{k,m}\}$ as columns. Compute its chi-squared statistic and
    use (\ref{eq:wh}) to transform it to a 1-d.f. chi-squared value
    $W_I(X_1,X_2)$.
  \end{enumerate}
\end{procedure}

To control the frequency with which interaction tests are carried out,
we put a Bonferroni-corrected significance threshold on each set of
tests and carry out the interaction tests only if all main effects are
not significant. Let $\chi^2_{\nu,\alpha}$ denote the upper-$\alpha$
quantile of the chi-squared distribution with $\nu$ d.f.  The~algorithm
for variable selection can now be stated as follows.

\begin{algorithm}[(Variable selection)] \label{alg:select0}
Let
  $K > 1$ be the number of nonconstant predictor variables in the
  node. Define $\alpha = 0.05/K$ and $\beta = 0.05/\break \{K(K-1)\}$:
  \begin{enumerate}[3.]
  \item Use Procedure~\ref{proc:main} to find $W_M(X_i)$ for
    $i = 1, 2, \ldots, K$.
  \item If $\max_i W_M(X_i) > \chi^2_{1,\alpha}$, select the variable
    with the largest value of $W_M(X_i)$ and exit.
  \item Otherwise, use Procedure~\ref{proc:interact} to find
    $W_I(X_i,X_j)$ for each pair of predictor variables:
    \begin{enumerate}[(b)]
    \item[(a)] If $\max_{i\neq j} W_I(X_i,X_j) > \chi^2_{1,\beta}$, select
      the pair with the largest value of $W_I(X_i,X_j)$ and exit.
    \item[(b)] Otherwise, select the $X_i$ with the largest value of
      $W_M(X_i)$.
    \end{enumerate}
  \end{enumerate}
\end{algorithm}

\subsection{Split set selection}
\label{sec:splitset-select}

After Algorithm~\ref{alg:select0} selects a variable $X$ at node $t$,
we need to find a set of $X$-values to form the split $t = t_L \cup
t_R$, where $t_L$ and $t_R$ denote the left and right subnodes of
$t$. Let $p_L$ and $p_R$ denote the proportions of samples in $t$
going into $t_L$ and $t_R$, respectively.  We seek the split that
minimizes the weighted sum of Gini impurities
\begin{equation}
  p_L   g(t_L) + p_R   g(t_R).  \label{eq:2nodes}
\end{equation}
Let $n$ be the number of distinct values of $X$ in $t$.  If $X$ is
noncategorical, there are $(n-1)$ possible splits, with each split
point the mean of two consecutive order statistics.  If $X$ is
categorical, the number of possible splits is $(2^{n-1}-1)$, which
grows exponentially with $n$. If $J = 2$, however, we use the
following short-cut to reduce the search on the categorical variable
to $(n-1)$ splits. It is a special case of a more general result
proved in \cite{cart}, Section~9.4.

\begin{teorem} \label{th:1}
Suppose that $J=2$ and that $X$ is a
  categorical variable taking $n$ distinct values.  Let $r(a)$ denote
  the proportion of class~1 observations among those with $X = a$. Let
  $a_1 \prec a_2 \prec \cdots \prec a_n$ be an ordering of the $X$
  values such that $r(a_1) \leq r(a_2) \leq \cdots \leq r(a_n)$. Given
  any set $A$, let the observations be split into two groups $t_L =
  \{X \in A\}$ and $t_R = \{X \notin A\}$.  The~set $A$ that
  minimizes the function $p_L   g(t_L) + p_R   g(t_R)$ is $A_i =
  \{a_1, a_2, \ldots, a_i\}$ for some $i = 1, 2, \ldots, n-1$.
\end{teorem}

If $X$ is noncategorical, we carry out an exhaustive search for the
best split of the form $X \leq c$, with $c$ being the midpoint of two
consecutive order statistics. If $X$ is categorical, an exhaustive
search is done only if $n \leq 11$ or if $J = 2$ (using the above
shortcut). Otherwise, a restricted search is performed as follows:
\begin{enumerate}[2.]
\item If $J \leq 11$ and $n > 20$, divide the observations in $t$ into
  $n$ groups according to the categorical values of $X$ and find the
  class that minimizes the misclassification cost in each group.  Map
  $X$ to a new categorical variable $X'$ whose values are the
  minimizing class labels.  Carry out an exhaustive search for a split
  on $X'$ and re-express it in terms of $X$.
\item If $J > 11$ or $n \leq 20$, transform the categorical values
  into 0--1 dummy vectors and apply linear discriminant analysis (LDA)
  to them to find the largest discriminant coordinate $X''$.  Find the
  best split on $X''$ and then re-express it in terms of $X$. This
  technique is employed in Loh and Vanichsetakul \cite{fact}.
\end{enumerate}
The~complete details are given as Procedure~\ref{proc:splitset} in the
\hyperref[append]{Appendix}.

What to do if Algorithm~\ref{alg:select0} selects a pair of variables
$X_1$ and $X_2$, say? Then the split search is more complicated,
because the best split on $X_1$ may depend on how $X_2$ is
subsequently split. Similarly, the best split on $X_2$ should consider
the best subsequent splits on $X_1$.  Suppose that $t$ is split first
by one variable into $t_L$ and $t_R$, and that $t_L$ is split into
$t_{\mathit{LL}}$ and $t_{\mathit{LR}}$, and $t_R$ into $t_{\mathit{RL}}$ and $t_{\mathit{RR}}$, by the
other variable.  Let $p_{\mathit{LL}}$, $p_{\mathit{LR}}$, $p_{\mathit{RL}}$ and $p_{\mathit{RR}}$ denote
the proportions of samples in $t_{\mathit{LL}}$, $t_{\mathit{LR}}$, $t_{\mathit{RL}}$ and
$t_{\mathit{RR}}$, respectively. We select the split $t_L$ that minimizes
\begin{equation}
  p_{\mathit{LL}}  g(t_{\mathit{LL}}) + p_{\mathit{LR}}  g(t_{\mathit{LR}}) + p_{\mathit{RL}}  g(t_{\mathit{RL}}) + p_{\mathit{RR}}  g(t_{\mathit{RR}})
  \label{eq:4nodes}
\end{equation}
over $t_{\mathit{LL}}$, $t_{\mathit{LR}}$, $t_{\mathit{RL}}$ and $t_{\mathit{RR}}$.

Because this requires a two-level search, we restrict the number of
candidate splits to keep computation under control.  Let $m_0$ be a
user-specified number so that only splits yielding at least $m_0$
cases in each subnode are permitted.
For splits on noncategorical variables $X_k$, define $f =
\min(100N^{-1},1)$.  Let $\lfloor \cdot \rfloor$ be the greatest
integer function and let
\[ d = \min \{ \max(\lfloor fN(t) \rfloor,9), N(t)-2m_0+1\} \] be the
number of split points to be evaluated.  Clearly, $d < 100$.  Let
$x_k(i)$ denote the $i$th order statistic of $X_k$.  The~set of
restricted split points on $X_k$ are the members of the set
\begin{equation}
  \label{eq:Sk}
  S_k = \{x_k(i_1), x_k(i_2), \ldots, x_k(i_d)\},
\end{equation}
where $i_j=m_0+\lfloor j(N(t)-2m_0)/(d+1)\rfloor$ and $j = 1, 2,
\ldots, d$.  Technical details of the search, covering the cases where
0, 1 or 2 variables are categorical, are given in
Procedures~\ref{proc:2lin}--\ref{proc:splitcat2} in the \hyperref[append]{Appendix}.  The~entire split selection procedure can be stated as follows:

\begin{algorithm}[(Split selection)] \label{alg:simple}
  \begin{enumerate}[3.]
  \item Apply Algorithm~\ref{alg:select0} to the data in $t$ to select
    the split variable(s).
  \item If one variable is selected and it is categorical, use
    Procedure~\ref{proc:splitset} to split $t$ on that variable.
  \item If one variable is selected and it is noncategorical,
    search through all mid-points, $c$, of the ordered data values
    for the split $t_L = \{X \leq c\}$ that minimizes
    (\ref{eq:2nodes}).
  \item If two variables are selected, apply
    Procedure~\ref{proc:2lin}, \ref{proc:splitcat1} or
    \ref{proc:splitcat2} to split~$t$, depending on whether zero, one
    or two of the variables are categorical.
  \end{enumerate}
\end{algorithm}

\begin{figure}

\includegraphics{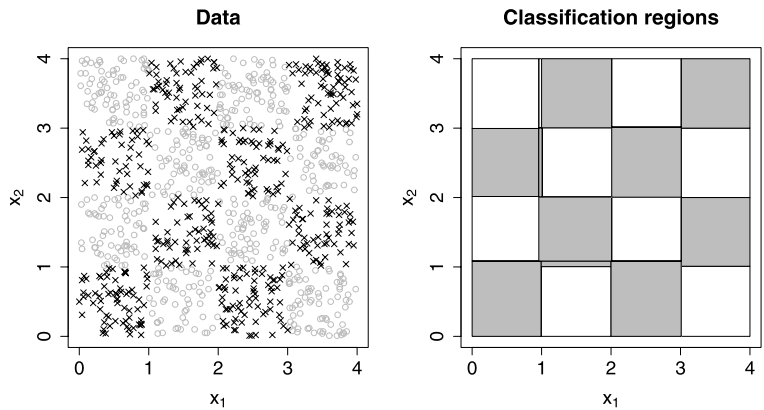}

  \caption{Data (left) and classification regions (right) for the
    \textit{S} method.}
  \label{fig:chess}
\end{figure}

The~power of this algorithm is best demonstrated by an artificial
example. We simulate one thousand observations, each randomly assigned
with probability 0.5 to one of two classes. Each class is uniformly
distributed in the alternating squares of a $4 \times 4$ chess board
in the $(X_1, X_2)$-plane. The~left side of Figure~\ref{fig:chess}
shows one realization, with 520 observations from class~1 and 480 from
class~2. We add eight independently and uniformly distributed noise
variables.  The~ideal classification tree should split on $X_1$ and
$X_2$ only and have 16 leaf nodes.  C4.5 and CTree yield trees with no
splits and hence misclassify 480 observations each.  RPART gives a
tree with 13 leaf nodes, but splits on $X_1$ or $X_2$ only three times
and misclassifies 347 observations.  QUEST misclassifies 27 with a
46-leaf tree.  CRUISE splits on $X_1$ or $X_2$ most of the time and
misclassifies 3, but because it does not perform two-level searches,
the tree is still large with 29 leaf nodes.  Our algorithm splits on
$X_1$ and $X_2$ exclusively and yields a 19-leaf tree that
misclassifies 4 observations; its classification regions are shown on
the right side of Figure~\ref{fig:chess}.

\section{Linear splits}
\label{sec:linsplits}
Although univariate splits (viz. splits on one variable at a time)
are best for interpretation, greater predictive accuracy can sometimes
be gained if splits on linear combinations of variables are allowed.
CART uses random search to find linear splits, while CRUISE and QUEST
use LDA; see also \cite{cline08,CK03,tabu03,YA05}.  Nonlinear splits
have been considered as well \cite{fan08,LDK05}.

\begin{figure}[b]

\includegraphics{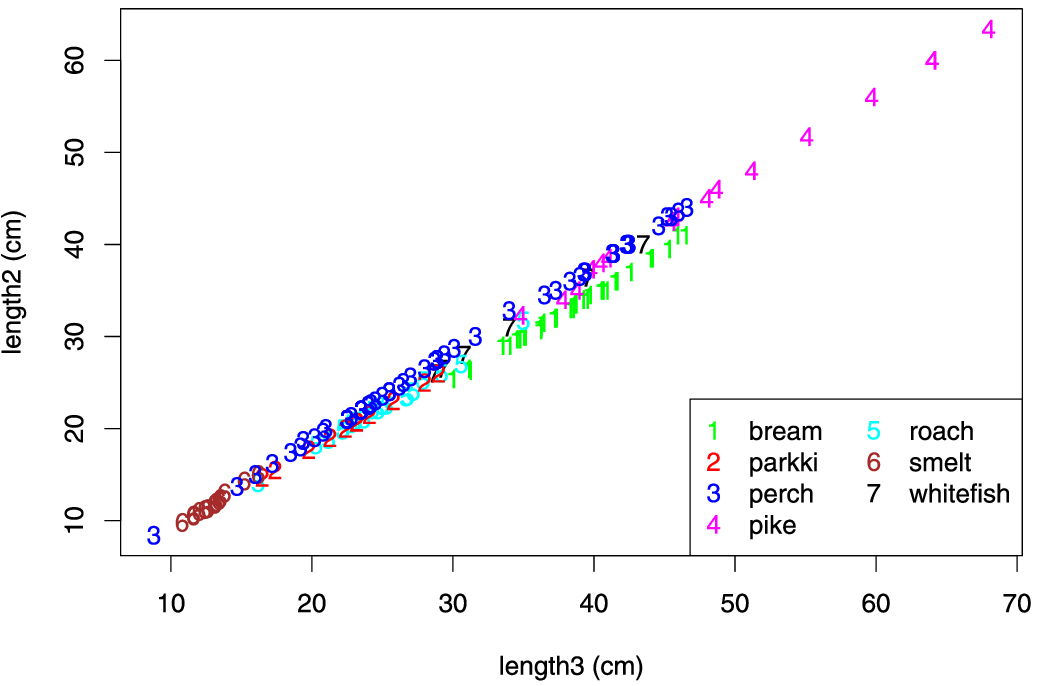}

  \caption{Plot of \texttt{length2} versus \texttt{length3} for fish data.}
  \label{fig:fish:lengths}
\end{figure}

To appreciate the necessity for linear splits, consider some data on
fish from Finland obtained from the \textit{Journal of Statistics
  Education} data archive
(\href{http://www.amstat.org/publications/jse/datasets/fishcatch.txt}{www.amstat.org/publications/jse/datasets/fishcatch.txt}).  See
\cite{SASUG} and \cite{cruise2v} for prior usage of these data.  There
are seven species (classes) in the sample of 159 fish. Their class
labels (in parentheses), counts and names are as follows: (1) 35 Bream, (2) 11
Parkki, (3) 56 Perch, (4) 17 Pike, (5) 20 Roach, (6) 14 Smelt and (7)
6 Whitefish.  Table~\ref{tab:fish} lists the seven predictor
variables.
The~data are challenging for univariate splits because the three
length variables are highly correlated.  For example,
Figure~\ref{fig:fish:lengths} shows that it is hard to separate the
classes using only univariate splits on \texttt{length2} and
\texttt{length3}.  A split along the direction of the data points,
however, can separate class~1 (bream) cleanly from the rest.

\begin{table}
  \caption{Predictor variables for fish data}
  \label{tab:fish}
      \begin{tabular*}{\textwidth}{@{\extracolsep{\fill}}ll@{}}
      \hline
      \texttt{weight}  & Weight of the fish (in grams) \\
      \texttt{length1} & Length from the nose to the beginning of the tail (in cm) \\
      \texttt{length2} & Length from the nose to the notch of the tail (in cm) \\
      \texttt{length3} & Length from the nose to the end of the tail (in cm) \\
      \texttt{height} & Maximal height as a percentage of \texttt{length3} \\
      \texttt{width} & Maximal width as a percentage of \texttt{length3} \\
      \texttt{sex}   & Male or female \\
      \hline
    \end{tabular*}
\end{table}

The~CRUISE 2V algorithm \cite{cruise2v} does well here because it
fits a linear discriminant model to a pair of noncategorical
variables in each node. We propose instead to keep the node models
simple, but use LDA to find splits on two variables at a time. The~restriction to two variables permits each split to be presented
graphically. It also reduces the impact of missing values---the more
variables are involved in a split, the fewer the number of complete
cases for its estimation.  To prevent outliers from having undue
effects on the estimation of the split direction, we trim them before
application of LDA in the following procedure.  As before, we use
chi-squared tests to select the variables for each linear split.

\begin{procedure} \label{proc:linear}
    Discriminant chi-squared statistic.  Let $X_1$ and $X_2$ be two
    noncategorical variables to be used in a linear split of node $t$:
  \begin{enumerate}[3.]
  \item For class $j$ ($j = 1, 2, \ldots, J$) and each $X_i$ ($i = 1,
    2$), compute the class mean $\bar{x}_{i,j}$ and class standard
    deviation $s_{i,j}$ of the samples in $t$.
  \item Let $S_j$ denote the set of class $j$ samples in $t$ lying in
    the rectangle $|X_i - \bar{x}_{i,j}| \leq 2s_{i,j}$ for $i = 1,
    2$.
  \item Find the larger linear discriminant coordinate $Z$ from the
    observations in $S_1 \cup \cdots \cup S_J$.
  \item Project the data in $t$ onto the $Z$-axis and compute their
    $Z$-values.
  \item Apply Procedure~\ref{proc:main} to the $Z$-values to find
    the one-d.f. chi-squared value $W_L(X_1,X_2)$.
  \end{enumerate}
\end{procedure}

Although linear splits are more powerful than univariate splits, it is
unnecessary to employ a linear split at every node.  We see from
Figure~\ref{fig:fish:lengths}, for example, that almost all the smelts
(class 6) can be separated from the other species (except for one
misclassified perch) by a univariate split on \texttt{length2} or on
\texttt{length3} at 14 cm.  Therefore, we should invoke linear splits
only if the main effect and interaction chi-squared tests are not
significant, and then only if the linear split itself is
significant. This differs from the linear split options in CART,
CRUISE and QUEST, which always split on linear combinations of all
the predictors. The~next algorithm replaces
Algorithms~\ref{alg:select0} and \ref{alg:simple}, if linear splits
are desired.

\begin{algorithm}[(Split selection with the
    linear split option)] \label{alg:select}
  Let $K > 0$ be the number of nonconstant
  predictor variables in node $t$ and let $K_1$ be the number among
  them that are noncategorical. Let $\alpha = 0.05/K$ and let $\beta
  = 0.05/\{K(K-1)\}$ if $K > 1$; otherwise, let $\beta = 0$.  Further,
  let $\gamma = 0.05/\{K_1(K_1-1)\}$ if $K_1 > 1$, and let it be 0
  otherwise:
  \begin{enumerate}[3.]
  \item If $K = 1$, define $W_M = \infty$ for the nonconstant
    variable.  Otherwise, perform Procedure~\ref{proc:main} to find
    $W_M(X_i)$ for $i = 1, 2, \ldots, K$.
  \item If $\max_i W_M(X_i) > \chi^2_{1,\alpha}$, let $X'$ be the
    variable with the largest value of $W_M(X_i)$: \label{step:2}
    \begin{enumerate}[(a)]
    \item[(a)] If $X'$ is categorical, split $t$ on $X'$ with
      Procedure~\ref{proc:splitset} and exit.
    \item[(b)] Otherwise, split $t$ at the midpoint of the ordered $X'$
      values that minimizes the sum of Gini impurities
      (\ref{eq:2nodes}) and exit.
    \end{enumerate}
  \item If $\max_i W_M(X_i) \leq \chi^2_{1,\alpha}$, use
    Procedure~\ref{proc:interact} to find $W_I(X_i,X_j)$ for each
    pair of predictor variables:
    \begin{enumerate}[(a)]
    \item[(a)] If $\max_{i\neq j} W_I(X_i,X_j) > \chi^2_{1,\beta}$, select
      the pair with the largest value of $W_I(X_i,X_j)$ and use
      Procedure~\ref{proc:2lin}, \ref{proc:splitcat1} or
      \ref{proc:splitcat2} to split $t$, depending on whether 0, 1 or
      2 variables are categorical, and exit.
    \item[(b)] If $K_1 = 1$, go to step~\ref{step:last}(c)ii below.
    \item[(c)] If $K_1 > 1$, use Procedure~\ref{proc:linear} to find
      $W_L(X_i,X_j)$ for each pair of noncategorical variables:
      \begin{enumerate}[ii.]
      \item[i.] If $\max_{i\neq j} W_L(X_i,X_j) > \chi^2_{1,\gamma}$,
        select the pair of variables with the largest
        $W_L(X_i,X_j)$-value, split $t$ on their larger discriminant
        coordinate, and exit.
      \item[ii.] Otherwise, let $X'$ be the variable with the largest value
        of $W_M(X_i)$ and go to step~\ref{step:2} above. \label{step:last}
      \end{enumerate}
    \end{enumerate}
  \end{enumerate}
\end{algorithm}

\begin{figure}

\includegraphics{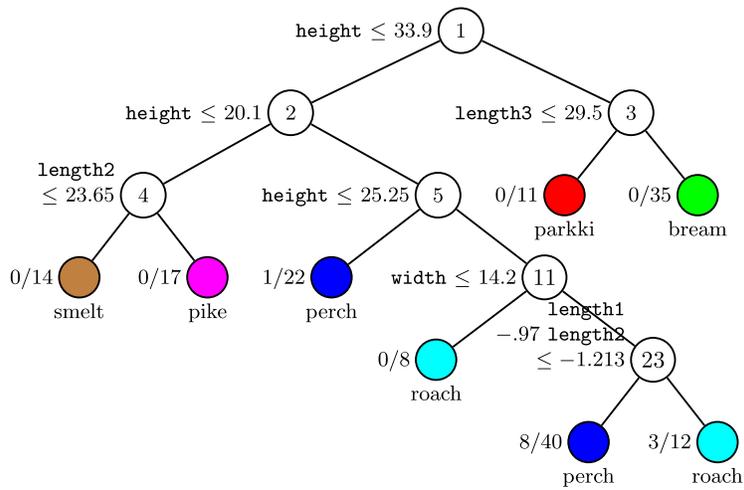}

  \caption{Tree model using the \texttt{S} method for the fish data.
    At each intermediate node, a case goes to the left child node if
    and only if the condition is satisfied. The~predicted class and
    number of errors divided by sample size are printed below and to
    the left of each leaf node.}
  \label{fig:fish:tree}
\end{figure}

\begin{figure}

\includegraphics{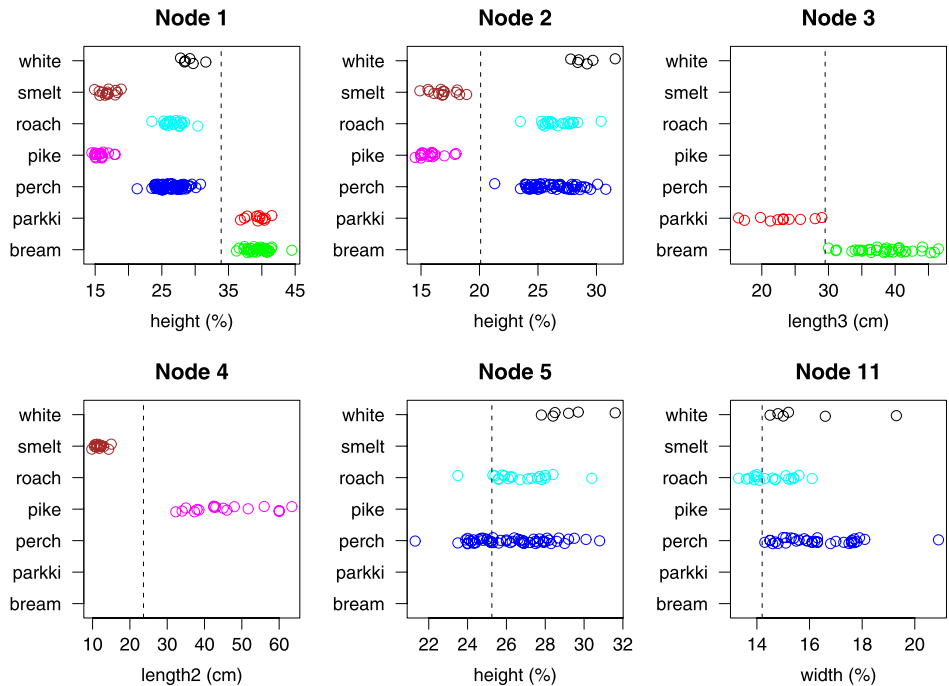}

  \caption{Plots of fish data in the intermediate nodes of the tree in
    Figure~\protect\ref{fig:fish:tree}, with dashed lines marking the
    splits. Points are vertically jittered.}
  \label{fig:fish:p1}
\end{figure}

\begin{figure}[b]

\includegraphics{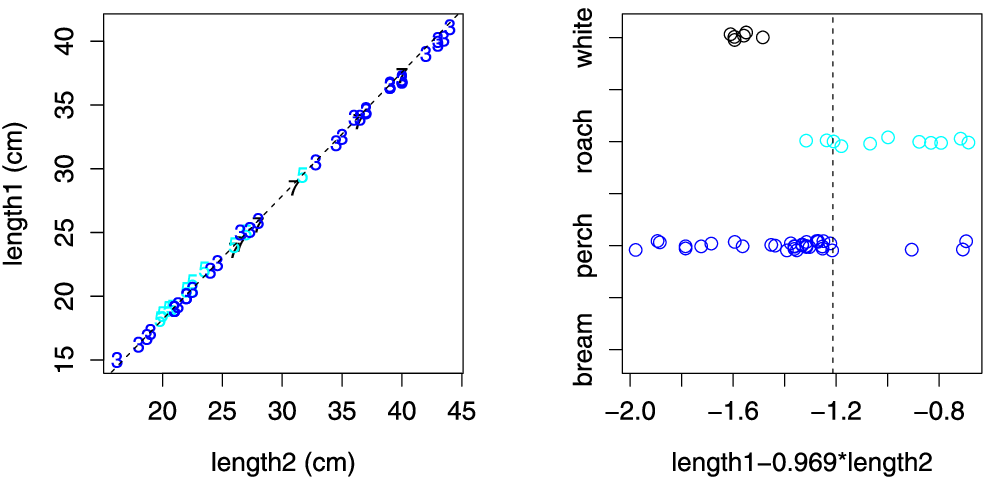}

  \caption{Plots of fish data in node~23 of the tree in
    Figure~\protect\ref{fig:fish:tree}, with dashed lines marking the
    split. The~left panel shows the data in terms of variables
    \texttt{length1} and \texttt{length2}; the right panel is a
    jittered plot of the same data in the linear discriminant
    direction.  The~plot symbols in the left panel are the same as
    those in Figure~\protect\ref{fig:fish:lengths}.}
  \label{fig:fish:p2}
\end{figure}

Figure~\ref{fig:fish:tree} shows the pruned tree for the fish data
using this algorithm. The~tree has univariate splits everywhere except
at node~23, where it uses a linear split on variables \texttt{length1}
and \texttt{length2}, and misclassifies 12 observations.
Figure~\ref{fig:fish:p1} displays jittered plots of the data in the
intermediate nodes with univariate splits.  Figure~\ref{fig:fish:p2}
shows two views of the data in the node with the linear split; the
left panel displays the data in terms of \texttt{length1} and
\texttt{length2}, and the right panel shows them in terms of the
linear discriminant coordinate.  Obviously, it would be impossible to
show the left panel if the linear split involves more than two
variables.

If linear splits are disallowed, the tree has 12 leaf nodes and
misclassifies 10 observations. The~CRUISE, QUEST, CTree and RPART
trees have 16, 5, 7 and 6 leaf nodes and misclassify 24, 26, 28 and
21 observations, respectively.  CRUISE 2V, which employs bivariate
linear discriminant leaf node models, yields a tree with 5 leaf nodes
and misclassifies 3.

The~variable definitions in Table~\ref{tab:fish} are taken directly
from the data source. If the length, width and height variables are
all expressed as percentages of \texttt{length3}, the resulting tree
has 6 leaf nodes with no linear splits and misclassifies 7
observations. Conversely, if all are expressed in cm instead of
percentages, the tree has 15 leaf nodes, employs 4 linear splits and 1
interaction split, and misclassifies 17.  Thus, transformations of the
variables can make a difference.  We note, further, that the error
rates are probably biased low, because they are estimated from the
same data set used to build the models.  Cross-validation estimates
are used to compare algorithms in Section~\ref{sec:comp}.

\section{Selection bias}
\label{sec:bias}

As mentioned earlier, it is important for an algorithm to be without
variable selection bias.  Unbiasedness here refers to the property
that the $X$ variables are selected with equal probability, when each
is independent of the class variable. To show that
Algorithm~\ref{alg:select} is practically unbiased, we report the
results of a simulation experiment with $N = 500$ and six $X$
variables. The~class variable takes two values with equal
probabilities and is independent of the $X$ variables.  We consider
two scenarios. In the \textit{independence} scenario, the $X$ variables
are mutually independent with the following distributions:
\begin{enumerate}[6.]
\item $X_1$ is categorical with $P(X_1 = 1) = P(X_2 = 2) = 1/2$;
\item $X_2$ is categorical with $P(X_2 = 1) = 1/6$, $P(X_2 = 2) =
  1/3$, and $P(X_2 = 3) = 1/2$;
\item $X_3$ is categorical taking six values with equal probability;
\item $X_4$ is chi-squared with one degree of freedom;
\item $X_5$ is normal;
\item $X_6$ is uniform on the unit interval.
\end{enumerate}
In the \textit{dependence} scenario, $X_1$ and $X_6$ are independent and
distributed as before, but $X_4$ and $X_5$ are bivariate normal with
correlation 0.7 and $X_2$ and $X_3$ have the joint distribution shown
in Table~\ref{tab:unbiased:dep}.

\begin{table}[b]
   \caption{Joint distribution of $X_2$ and $X_3$ in the dependence scenario}
  \label{tab:unbiased:dep}
  \begin{tabular*}{\textwidth}{@{\extracolsep{\fill}}lcccccc@{}}
  \hline
    & \multicolumn{6}{c@{}}{$\bolds{X_3}$} \\[-6pt]
     & \multicolumn{6}{c@{}}{\hrulefill} \\
    $\bolds{X_2}$ & \textbf{1} & \textbf{2} & \textbf{3} & \textbf{4} & \textbf{5} & \textbf{6} \\
    \hline
    1 & 1$/$12 & 1$/$12 & 1$/$24 & 1$/$24 & 1$/$24 & 1$/$24 \\
    2 & 1$/$24 & 1$/$24 & 1$/$12 & 1$/$12 & 1$/$24 & 1$/$24 \\
    3 & 1$/$24 & 1$/$24 & 1$/$24 & 1$/$24 & 1$/$12 & 1$/$12 \\
    \hline
  \end{tabular*}
\end{table}

\begin{table}
   \caption{Number of times (out of 10,000 simulation trials) that each variable is
    selected to split the root node. Variables $X_1$, $X_2$ and $X_3$ are
    categorical and can only appear in a univariate split; variables $X_4$, $X_5$
     and $X_6$ can appear in a univariate or a linear split}
  \label{tab:unbiased:counts}
  \begin{tabular*}{\textwidth}{@{\extracolsep{\fill}}lccccccccc@{}}
  \hline
    & \multicolumn{6}{c}{\textbf{Univariate splits}} & \multicolumn{3}{c@{}}{\textbf{Linear splits}} \\[-6pt]
     & \multicolumn{6}{c}{\hrulefill} & \multicolumn{3}{c@{}}{\hrulefill} \\
    \textbf{Scenario} & $\bolds{X_1}$ & $\bolds{X_2}$ & $\bolds{X_3}$ & $\bolds{X_4}$ & $\bolds{X_5}$ & $\bolds{X_6}$ & $\bolds{X_4}$ & $\bolds{X_5}$ & $\bolds{X_6}$ \\
    \hline
    Independence & 1713 & 1656 & 1620 & 1567 & 1492 & 1533 & 302 & 310 & 226 \\
    Dependence & 1684 & 1643 & 1601 & 1544 & 1564 & 1503 & 356 & 342 & 224 \\
    \hline
  \end{tabular*}
\end{table}

\begin{table}[b]
    \caption{Probabilities of variable selection for a null model estimated by
    simulation. The~simulation standard error is 0.0037. If the method is unbiased,
    the probabilities should be all equal to 0.1667}
  \label{tab:unbiased:rates}
  \begin{tabular*}{\textwidth}{@{\extracolsep{\fill}}lcccccc@{}}
  \hline
    & $\bolds{X_1}$ & $\bolds{X_2}$ & $\bolds{X_3}$ & $\bolds{X_4}$ & $\bolds{X_5}$ & $\bolds{X_6}$ \\
    \hline
    Independent & 0.1713 & 0.1656 & 0.1620 & 0.1718 & 0.1647 & 0.1646 \\
    Dependent & 0.1684 & 0.1643 & 0.1601 & 0.1722 & 0.1735 & 0.1615 \\
    \hline
  \end{tabular*}
\end{table}

Table~\ref{tab:unbiased:counts} shows the number of times each
variable is chosen over 10,000 simulation trials.  Among univariate
splits, there is a slightly lower probability that a noncategorical
variable is selected (most likely due to discretization of the
continuous variables or the Wilson--Hilferty approximation), but it is
offset by the probability that such variables are selected through
linear splits.  Because two $X$ variables are involved in a linear
split, each one is double-counted in the three columns on the right
side of Table~\ref{tab:unbiased:counts}. Therefore, to estimate the
overall selection probability for each variable, we halve these counts
before adding them to the corresponding univariate split counts. The~results, shown in Table~\ref{tab:unbiased:rates}, are all within two
simulation standard errors of $1/6$ (the required value for
unbiasedness).

\section{Kernel and nearest-neighbor node models}
\label{sec:knn}
So far, we have tried to make the tree structure more parsimonious and
precise by controlling selection bias and improving the discriminatory
power of the splits.  Another way to reduce the size of the tree
structure is to fit nontrivial models to the nodes.  Kim and Loh
\cite{cruise2v} and Gama \cite{Gama04}, for example, use linear
discriminant models.  Although effective in improving predictive
accuracy, linear discriminant models are not as flexible as
nonparametric ones and may not simplify the tree structure as much. We
propose to use kernel and nearest-neighbor models instead. To allow
the fits to be depicted graphically, we again restrict each model to
use at most two variables.  Further, to save computation time, we use
Algorithm~\ref{alg:simple} to find the split variables, skipping the
linear splits.  Buttrey and Karo \cite{BK02} also fit nearest-neighbor
models, but they do this only after a standard classification tree is
constructed.  As a result, their tree structure is unchanged by model
fitting. Since we fit a model to each node as the tree is grown, we
should get more compact trees.

First, consider kernel discrimination, which is basically maximum
likelihood with a kernel density estimate for each class in a node.
If the selected $X$ variable is categorical, its class density
estimate is just the sample class density function.  If $X$ is
noncategorical, we use a Gaussian kernel density estimate.  Let $s$
and $r$ denote the standard deviation and the inter-quartile range,
respectively, of a sample of observations, $x_1, x_2, \ldots, x_n$,
from $X$.  The~kernel density estimate is $\hat{f}(x) = (nh)^{-1}
\sum_{i=1}^n \phi\{(x-x_i)/h\}$, where $\phi$ is the standard normal
density function and $h$ is the bandwidth. The~following formula,
adapted from Stata \cite{stata},
\begin{equation}\label{eq:bandwidth}
  h = \cases{
          2.5\min(s, 0.7413r)n^{-1/5}, &\quad  if $r > 0$, \vspace*{2pt}\cr
      2.5sn^{-1/5}, &\quad  otherwise,}
\end{equation}
is used in the calculations reported here.  This bandwidth is more
than twice as wide as the asymptotically optimal value usually
recommended for density estimation; Ghosh and Chaudhuri \cite{GCS06}
find that the best bandwidth for discrimination is often much larger
than that for density estimation.  For our purposes, asymptotic
formulas cannot be taken too seriously, because the node sample size
decreases with splitting.

If a pair of noncategorical variables is selected, we fit a bivariate
kernel density to the pair for each class. If the split is due to one
categorical and one noncategorical variable, we fit a kernel density
estimate to the noncategorical variable for each class and each value
of the categorical variable, using an average bandwidth that depends
only on the class.  Averaging smoothes out the effects of small or
highly unbalanced sample sizes.  The~details are given in the next
algorithm.

\begin{algorithm}[(Kernel models)] \label{alg:kernel}
   Let $Y$ denote the class variable and apply
  Algorithm~\ref{alg:simple} to find one or two variables to split a
  node $t$:
  \begin{enumerate}[2.]
  \item If the split is due to a main effect chi-squared statistic
    (Procedure~\ref{proc:main}), let $X$ be the selected
    variable. Fit a kernel density estimate to $X$ for each class in
    $t$ using bandwidth (\ref{eq:bandwidth}) with $n = N(t)$.
  \item If the split is due to an interaction chi-squared statistic
    (Procedure~\ref{proc:interact}), let $X_1$ and $X_2$ be the
    selected variables. Fit a bivariate density estimate to $(X_1,
    X_2)$ for each class in $t$ as follows:
    \begin{enumerate}[(a)]
    \item[(a)] If $X_1$ and $X_2$ are categorical, use their sample class
      joint density.
    \item[(b)] If $X_1$ is categorical and $X_2$ is noncategorical, then
      for each combination of $(X_1, Y)$ values present in $t$, find a
      bandwidth $h(Y,X_1)$ using (\ref{eq:bandwidth}).  Let
      $\bar{h}(Y)$ be the average of $h(Y,X_1)$. For each value of
      $X_1$ and $Y$, find a kernel density estimate for $X_2$ with
      $\bar{h}(Y)$ as bandwidth.
    \item[(c)] If both variables are noncategorical, fit a bivariate
      Gaussian kernel density estimate to each class with correlation
      equal to the class sample correlation and $n$ equal to the class
      sample size in (\ref{eq:bandwidth}).
    \end{enumerate}
  \end{enumerate}
  The~predicted class is the one with the largest estimated
  density.
\end{algorithm}

We use the same idea for nearest-neighbor models. For noncategorical
variables, the number of nearest neighbors, $k$, is given by the formula
\begin{equation}
  \label{eq:knn}
  k = \max(3, \lceil \log n \rceil),
\end{equation}
where $n$ is the number of observations and $\lceil x \rceil$ denotes
the smallest integer greater than or equal to $x$.  We require $k$ to
be no less than 3 to lessen the chance of ties. The~full details are
given next.

\begin{algorithm}[(Nearest neighbor models)] \label{alg:nn}
 Let
  $\hat{Y}$ denote the predicted value of $Y$. Use
  Algorithm~\ref{alg:simple} to find one or two variables to split a
  node $t$:
  \begin{enumerate}[2.]
  \item If the split is due to a main effect chi-squared statistic
    (Procedure~\ref{proc:main}), let $X$ be the selected variable:
    \begin{enumerate}[(a)]
    \item[(a)] If $X$ is categorical, then $\hat{Y}$ is the highest
      probability class among the observations in $t$ with the same
      $X$ value as the one to be classified.
    \item[(b)] If $X$ is noncategorical, then $\hat{Y}$ is given by the
      $k$-nearest neighbor classifier based on $X$ with $n = N(t)$ in
      (\ref{eq:knn}).
    \end{enumerate}
  \item If the split is due to an interaction chi-squared statistic
    (Procedure~\ref{proc:interact}), let $X_1$ and $X_2$ be the
    selected variables:
    \begin{enumerate}[(a)]
    \item[(a)] If both variables are categorical, then $\hat{Y}$ is the
      highest probability class among the observations in $t$ with the
      same $(X_1, X_2)$ values as the one to be classified.
    \item[(b)] If $X_1$ is categorical and $X_2$ is noncategorical, then
      $\hat{Y}$ is given by the \mbox{$k$-nearest} neighbor classifier based
      on $X_2$ applied to the set $S$ of observations in $t$ that have
      the same $X_1$ value as the one to be classified, with $n$ being
      the size of $S$ in (\ref{eq:knn}).
    \item[(c)] If both variables are noncategorical, then $\hat{Y}$ is
      given by the bivariate \mbox{$k$-nearest} neighbor classifier based on
      $X_1$ and $X_2$ with the Mahalanobis distance and $n = N(t)$ in
      (\ref{eq:knn}).
    \end{enumerate}
  \end{enumerate}
\end{algorithm}

\begin{figure}[b]

\includegraphics{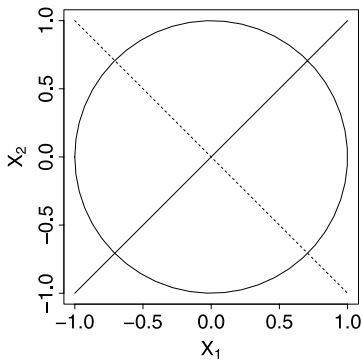}

  \caption{Three-class problem, with class 1 uniformly distributed on
    the circle and classes~2 and 3 each uniformly distributed on its own
    diagonal line.}
  \label{fig:xc}
\end{figure}

Figure~\ref{fig:xc} shows an artificial example that is very
challenging for many algorithms.  There are 300 observations, with 100
from each of three classes and eight predictor variables. Class~1 is
uniformly distributed on the unit circle in the $(X_1, X_2)$-plane.
Class~2 is uniformly distributed on the line $X_1-X_2=0$, and class~3
on the line $X_1+X_2=0$, with $|X_1|<1$ and $|X_2|<1$. Variables
$X_3$, $X_4$ and~$X_5$ are uniformly distributed on the unit interval
and $X_6$, $X_7$ and~$X_8$ are categorical, each taking 21
equi-probable values.  The~$X$ variables are mutually independent.

QUEST with linear splits gives a 38-leaf tree that misclassifies 3
samples.  CRUISE with LDA models gives a 17-leaf tree that
misclassifies 10 samples.  RPART, QUEST with univariate splits and
our simple node method are about equal, misclassifying 85, 81 and 75
samples, respectively.  C4.5 and CTree are the worst, with 134 and 200
misclassified and 84 and 1 leaf nodes, respectively.  In contrast, our
kernel and nearest-neighbor methods yield trees with no splits after
pruning and misclassify 2 and 8 training samples, respectively.

\begin{table}[b]
  \caption{Algorithms and plot symbols for the comparison experiment}
  \label{tab:names}
  \begin{tabular*}{\textwidth}{@{\extracolsep{\fill}}ll@{}}
  \hline
    \texttt{C45} & C4.5 \\
    \texttt{C2d} & CRUISE with interaction detection and simple node models \\
    \texttt{C2v} & CRUISE with interaction detection and linear discriminant node models \\
    \texttt{Qu} & QUEST with univariate splits \\
    \texttt{Ql} & QUEST with linear splits \\
    \texttt{Rp} & RPART \\
    \texttt{Ct} & CTree \\
    \texttt{S} & Proposed method with simple node models (Algorithm~\ref{alg:select})\\
    \texttt{K} & Proposed method with kernel node models
    (Algorithm~\ref{alg:kernel}) \\
    \texttt{N} & Proposed method with nearest-neighbor node models
    (Algorithm~\ref{alg:nn}) \\
    \hline
  \end{tabular*}
\end{table}

\section{Comparison on forty-six datasets}
\label{sec:comp}

The~error rates discussed so far are biased low because they are
computed from the same data that are used to construct the tree
models.  To obtain a better indication of relative predictive
accuracy, we compare the algorithms listed in Table~\ref{tab:names} on
forty-two real and four artificial data sets using ten-fold
cross-validation.  Each data set is randomly divided into ten roughly
equal parts and one-tenth is set aside in turn as a test set to
estimate the predictive accuracy of the model constructed from the
other nine-tenths of the data. The~cross-validation estimate of error
is the average of the ten estimates. Equal misclassification costs are
used throughout.  As elsewhere in this article, the trees are pruned
to have minimum cross-validation estimate of misclassification cost.
All other parameter values are set at their respective defaults.

The~four artificial data sets are those in Figures~\ref{fig:chess}
(\texttt{int}) and \ref{fig:xc} (\texttt{cl3}), and the digit
(\texttt{led}) and waveform (\texttt{wav}) data in \cite{cart}.
Sample size ranges from 97 to~45,222; number of classes from 2 to 11;
number of categorical variables from 0 to 60; number of
noncategorical variables from 0 to 69; and maximum number of
categories among categorical variables from 0 to 41. We use the
notations ``\texttt{S},'' ``\texttt{K}'' and ``\texttt{N}'' to refer
to our proposed method with simple (i.e., constant), kernel and
nearest-neighbor node models, respectively.  \texttt{S} employs linear
splits, but \texttt{K} and \texttt{N} do not.

Eleven data sets have missing values. In the \texttt{S}, \texttt{N}
and \texttt{K} algorithms, missing values in a categorical variable
are assigned to their own separate ``missing'' category. Observations
with missing values in a noncategorical split variable are always
sent to the left node. If there are enough such cases, our algorithms
will consider a split on missingness as one of the candidate splits.
Bandwidths are computed from cases with nonmissing values in the
selected variables. Kernel and nearest-neighbor model classifications
are applied only to observations with nonmissing values in the
selected variables. Observations with missing values in the selected
variables are classified with the majority rule.

\begin{figure}

\includegraphics{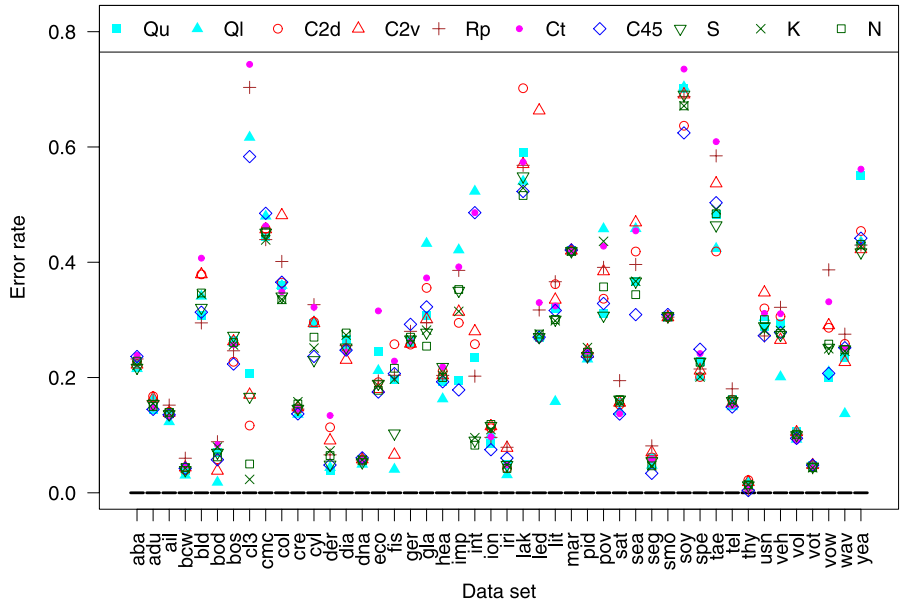}

  \caption{Error rates of algorithms by data set.}
  \label{fig:error}
\end{figure}

\begin{figure}

\includegraphics{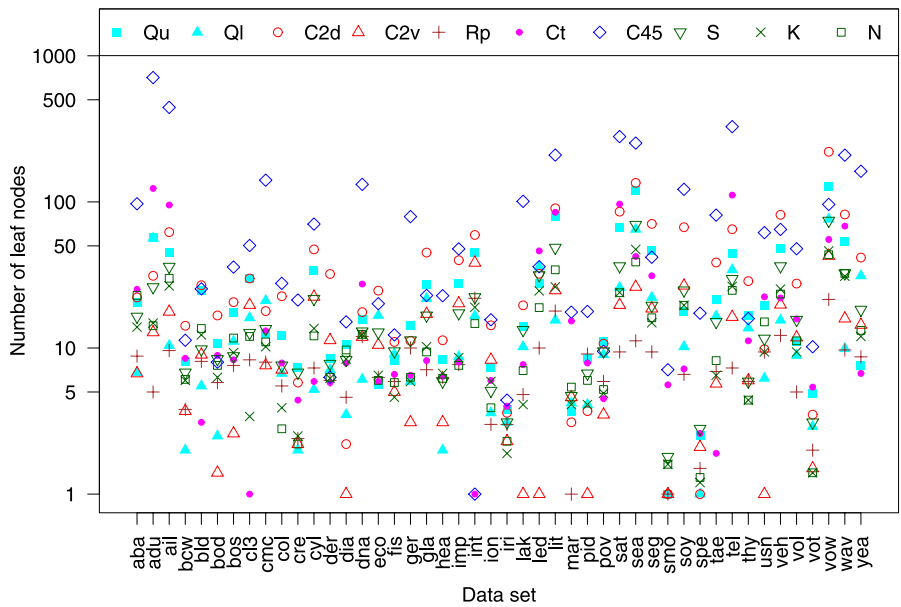}

  \caption{Numbers of leaf nodes by data set.}
  \label{fig:nodes}
\end{figure}

Figure~\ref{fig:error} graphs the errors rates of the ten algorithms
for each data set.  Despite the large range of the error rates (from
near 0 to about 0.7), the algorithms have very similar accuracy for
about half of the data sets.  The~most obvious exceptions are the
artificial data sets \texttt{int} and \texttt{cl3}, where our
\texttt{K}, \texttt{N} and \texttt{S} algorithms have a superior
edge; algorithms not designed to detect interactions pay a steep price
here. Two other examples showing interesting patterns are the fish
(\texttt{fis}) data used in Section~\ref{sec:linsplits} and the
\texttt{bod} data set \cite{body03}, where body measurements are used
to classify gender.  For these two data sets, algorithms that employ
LDA techniques (\texttt{C2v}, \texttt{Ql} and \texttt{S}) are more
accurate.  \texttt{Ql} seems to be either best or worst for a majority
of the data sets.

\begin{figure}[b]

\includegraphics{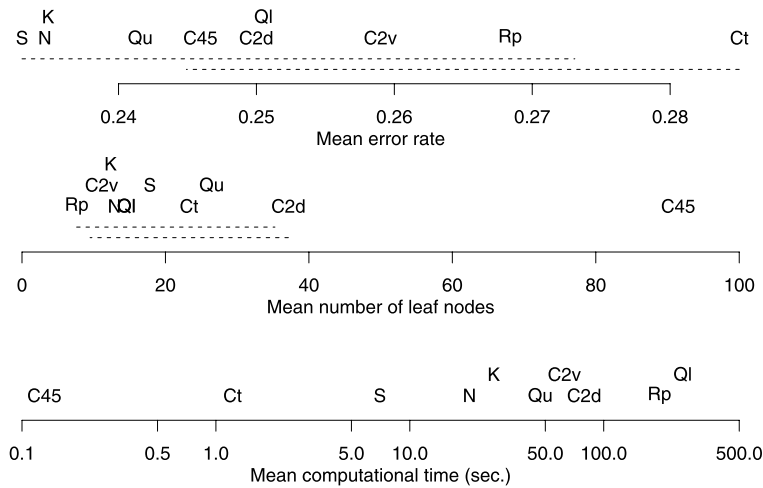}

  \caption{Means of error rates, numbers of leaf nodes and
    computational times; for the top two plots, dashed lines join
    algorithms that do not differ significantly at the 95\%
    simultaneous level of confidence.}
  \label{fig:tukey}
\end{figure}

Figure~\ref{fig:nodes} shows the corresponding results for the sizes
of the trees in terms of their numbers of leaf nodes.  \texttt{C45}
tends to produce the largest trees, while \texttt{C2v} and \texttt{Rp}
often give the shortest.

Figure~\ref{fig:tukey} shows arithmetic means (over the 46 data sets)
of the error rates and numbers of leaf nodes for each algorithm, with
corresponding 95\% Tukey simultaneous confidence intervals.  The~latter are obtained by fitting two-factor mixed models to the error
rates and number of leaf nodes separately, using algorithm as fixed
factor, data set as random factor, and their interaction as error
term.  \texttt{S} and \texttt{Ct} have the smallest and largest,
respectively, mean error rates. The~confidence intervals show that the
mean error rate of \texttt{Ct} differs significantly from those of
\texttt{S}, \texttt{N}, \texttt{K} and \texttt{Qu}; other differences
are not significant.  As for mean number of leaf nodes, \texttt{C45}
is significantly different from the others, as is \texttt{Rp} from
\texttt{C2d}.

Figure~\ref{fig:tukey} also shows the mean computational times, on a
2.66Ghz quad-core Linux PC with 8Gb memory.  Because the algorithms
are implemented in different computer languages (\texttt{Rp} and
\texttt{Ct} in R, \texttt{C45} in C and the others in Fortran), the
results compare execution times rather than number of computer
operations.  Further, the mean time for \texttt{Rp} is dominated by
two data sets each having six classes and categorical variables with
many categorical levels. It is well known that the computational time
of CART, upon which \texttt{Rp} is based, increases exponentially with
the number of categorical levels when the number of classes exceeds
two.

\begin{figure}

\includegraphics{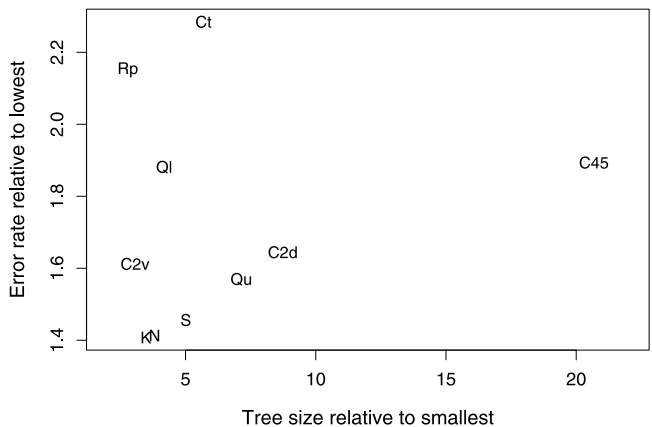}

  \caption{Mean of error rates relative to the lowest (for each data
    set) versus mean of tree size relative to that of the smallest
    tree; tree size is measured by the number of leaf nodes.}
  \label{fig:resultplots}
\end{figure}

Another way to account for differences among data sets is to compare
the ratio of the error rate of each algorithm to the lowest error rate
among the ten algorithms within each data set. We call this ratio the
``error rate relative to the lowest'' for the particular algorithm and
data set.  The~mean of these ratios for an algorithm over the data
sets yields an overall measure of its relative inaccuracy. Applying
the same procedure to the number of leaf nodes gives an overall
measure of relative tree size for each algorithm.
Figure~\ref{fig:resultplots} gives a plot of these two measures.  The~best algorithms are those in the bottom left corner of the plot:
\texttt{K}, \texttt{N}, \texttt{S} and \texttt{C2v}. They are
relatively most accurate and they yield relatively small trees.

\section{Tree ensembles}
\label{sec:ensembles}
A tree ensemble classifier uses the majority vote from a collection of
tree models to predict the class of an observation. \textit{Bagging}
\cite{bagging96} creates the ensemble by using bootstrap samples of
the training data to construct the trees.  Random Forest
(\texttt{RF}), which is based on CART and employs 500 trees, goes
beyond bagging by splitting each node on a random subset of $\sqrt{K}$
variables ($K$ being the total number of variables) and not pruning
the trees. Because it is practically impossible to interpret so many
trees, ensemble classifiers are typically used for prediction only.

\begin{figure}[b]

\includegraphics{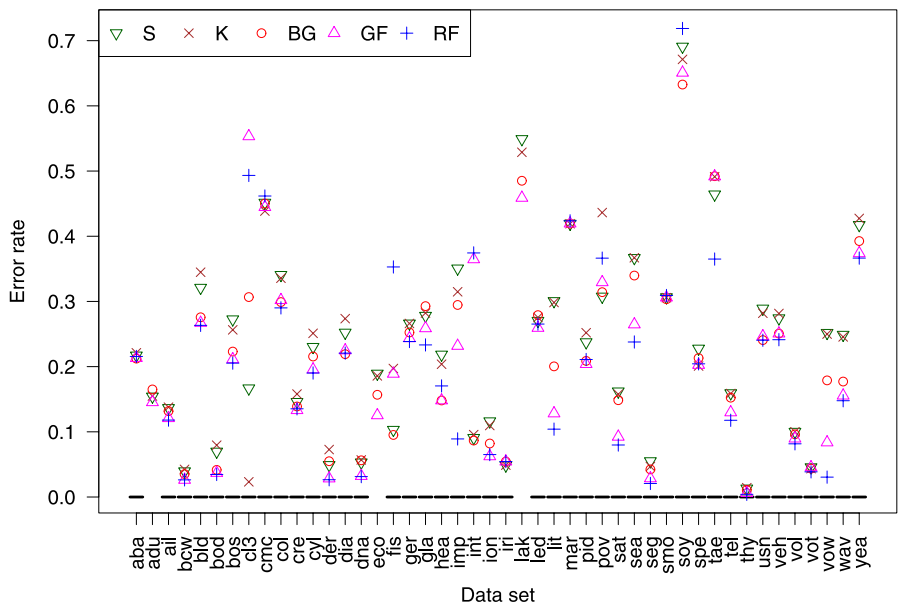}

  \caption{Single tree models (\texttt{S} and \texttt{K}) versus
  ensemble models (\texttt{BG}, \texttt{GF} and \texttt{RF}).}
  \label{fig:ensemble}
\end{figure}

To see how much an ensemble can affect the predictive accuracy of the
single-tree models proposed here, we use the bagging and \texttt{RF}
ideas to create two new ensemble classifiers. The~first, called
\textit{bagged GUIDE} (\texttt{BG}), is a collection of 100 pruned
trees, each constructed using the \texttt{S} method from bootstrap
samples. The~second ensemble classifier, called \textit{GUIDE forest}
(\texttt{GF}), consists of 500 unpruned trees constructed by the
\texttt{S} method without interaction and linear splits.  As in
\texttt{RF}, \texttt{GF} uses a random subset of $\sqrt{K}$ variables
to split each node.

Figure~\ref{fig:ensemble} shows the error rates of \texttt{BG},
\texttt{GF} and \texttt{RF} compared with those of the \texttt{S} and
\texttt{K} methods for each data set.  The~R package of Liaw and
Wiener \cite{rf}, which we use for \texttt{RF}, is not applicable if
the data set has predictor variables with more than 32 categorical
levels or if the test sample contains class labels that are not
present among the training samples. The~first condition occurs in the
\texttt{adu} and \texttt{lak} data sets, which have categorical
variables with 41 and 35 levels, respectively, and the second
condition occurs with the \texttt{eco} data set, which has 8 classes
and a total sample size of 336. Table~\ref{tab:ensemble:means} gives
the mean error rates with these three data sets excluded. All three
ensemble methods have lower means than the single-tree methods, but
the differences are fairly small on average.
Figure~\ref{fig:ensemble} shows that although \texttt{RF} is best for
many data sets, it performs particularly poorly compared to \texttt{S}
and \texttt{K} for three data sets: \texttt{cl3}
(Figure~\ref{fig:xc}), \texttt{fis} (fish data in
Section~\ref{sec:linsplits}) and \texttt{int}
(Figure~\ref{fig:chess})---data sets that have strong linear or
interaction effects. \texttt{GF} shares the same difficulties with
\texttt{RF}, but \texttt{BG} does not, because the latter allows
linear and interaction splits.

\begin{table}
  \caption{Mean error rates over the 46 data sets where \texttt{RF} is applicable.
    Differences are not statistically~significant}
  \label{tab:ensemble:means}
  \begin{tabular*}{\textwidth}{@{\extracolsep{\fill}}lccccc@{}}
  \hline
    \textbf{Algorithm} & \textbf{\texttt{S}} & \textbf{\texttt{K}} & \textbf{\texttt{BG}} & \textbf{\texttt{GF}} & \textbf{\texttt{RF}} \\
    \hline
    Error rate & 0.228 & 0.231 & 0.212 & 0.212 & 0.206 \\
    \hline
  \end{tabular*}
\end{table}

\section{Conclusion}
\label{sec:conc}

Improving the prediction accuracy of a tree and the precision of its
splits is a balancing act. On the one hand, we must refrain from
searching too greedily for splits, as the resulting selection bias can
cause irrelevant variables to be selected. On the other hand, we
should search hard enough to avoid overlooking good splits hidden
behind interactions or linear relationships. We solve this problem by
using three groups of significance tests, with increasing complexity
of effects and decreasing order of priority. The~first group of tests
for main effects is always carried out. The~second group, which tests
for interactions, is performed only if no main effect test is
significant.  The~third group, which tests for linear structure, is
performed only if no test in the first two groups is significant. A
Bonferroni correction controls the significance level of each
group. In addition, if an interaction test is significant, the split
is found by a two-level search on the pair of interacting variables.

If greater reduction in the size of the tree structure is desired, we
can fit a kernel or a nearest-neighbor model to each node.  Owing to
the flexibility of these models, we dispense with linear splits in
such situations.  We showed by an example that when there are weak
main effects but strong two-factor interaction effects, classification
trees constructed with these models can achieve substantial gains in
accuracy and tree compactness. They require more computation than
trees with simple node models, but the empirical evidence indicates
that their prediction accuracy remains high even for ordinary data
sets.

We also investigated the effect of tree ensembles on predictive
accuracy.  Although Random Forest quite often yields higher accuracy
than the single-tree models \texttt{S} and \texttt{K}, the average
increase is only about 10\% for the 43 data sets in the study. Much
depends on the complexity of the data. If there are strong interaction
or linear effects, the single-tree algorithms proposed here can be
substantially more accurate than Random Forest.  Further, the choice
of the single-tree algorithm used in the ensemble matters.

All the proposed algorithms are implemented in version~8 of GUIDE,
which may be obtained from \href{http://www.stat.wisc.edu/\textasciitilde loh/guide.html}{www.stat.wisc.edu/\textasciitilde loh/guide.html} for
the Linux, Macintosh and Windows operating systems.

\begin{appendix}\label{append}
\section*{Appendix}

\begin{procedure} \label{proc:splitset}
   Split set selection for a categorical variable $X$.  Let $\{a_1,
  a_2, \ldots, a_n\}$ be the set of distinct values of $X$ in node
  $t$:
  \begin{enumerate}[3.]
  \item If $J = 2$ or $n \leq 11$, search all subsets $S$ to find a
    split of the form $t_L = \{X \in S\}$ that minimizes
    (\ref{eq:2nodes}).
  \item Else if $J \leq 11$ and $n > 20$, for each $i=1, 2, \ldots,
    n$, let $j_i$ be the class that minimizes the misclassification
    cost for the observations in $t \cap \{X = a_i\}$. Define the new
    categorical variable $X' = \sum_i j_i   I(X = a_i)$ and search
    all subsets $S'$ to find a split of the form $\{X' \in S'\}$ that
    minimizes (\ref{eq:2nodes}). Re-express the selected split in
    terms of $X$ as $t_L = \{X \in S\}$.
  \item Else use LDA as follows:
    \begin{enumerate}[(c)]
    \item[(a)] Convert $X$ into a vector of dummy variables $(u_1, u_2,
      \ldots, u_n)$, where $u_i = 1$ if $X = a_i$ and $u_i = 0$
      otherwise.
    \item[(b)] Obtain the covariance matrix of the $u$-vectors and find the
      eigenvectors associated with the positive eigenvalues.  Project
      the $u$-vectors onto the space spanned by these eigenvectors.
    \item[(c)] Apply LDA to the projected
      $u$-vectors to find the largest discriminant coordinate $v =
      \sum_i c_i u_i$.
    \item[(d)] Let $v_{(1)} < v_{(2)} < \cdots$ denote the (at most $n$)
      sorted $v$-values.
    \item[(e)] Find the split $t_L = \{v \leq v_{(m)}\}$ that
      minimizes~(\ref{eq:2nodes}).
    \item[(f)] Re-express the split as $t_L = \{X \in S\}$.
    \end{enumerate}
  \end{enumerate}
 \end{procedure}

\begin{procedure} \label{proc:2lin}
     Split selection between two noncategorical variables $X_1$ and
    $X_2$.  Let $S_k$ ($k = 1, 2$) be defined as in (\ref{eq:Sk}):
  \begin{enumerate}[3.]
  \item Split $t$ first on $X_1$ and then on $X_2$ as follows. Given
    numbers $c$, $d$ and $e$, let $t_L = \{ X_1 \leq c\}$, $t_R = \{
    X_1 > c\}$, $t_{\mathit{LL}} = t_L \cap \{X_2 \leq d\}$, $t_{\mathit{LR}} = t_L \cap
    \{X_2 > d\}$, $t_{\mathit{RL}} = t_R \cap \{X_2 \leq e\}$, and $t_{\mathit{RR}} =
    t_R \cap \{X_2 > e\}$. Search over all $c \in S_1$ and $d, e \in
    S_2$ to find the best $c = c_1$ that minimizes~(\ref{eq:4nodes}).
  \item Exchange the roles of $X_1$ and $X_2$ in the preceding step to
    find the best split $\{X_2 \leq c_2\}$ with $c_2 \in S_2$.
  \item If the minimum value of~(\ref{eq:4nodes}) from the split
    $\{X_1 \leq c_1\}$ is less than that from $\{X_2 \leq c_2\}$,
    select the former.  Otherwise, select the latter.
  \end{enumerate}
\end{procedure}

\begin{procedure} \label{proc:splitcat1}
  Split selection between noncategorical $X_1$ and categorical~$X_2$:
  \begin{enumerate}[3.]
  \item First find a split of $t$ on $X_1$ as follows.  Let $t_L = \{
    X_1 \leq c\}$ and $t_R = \{ X_1 > c\}$, where $c \in S_1$ and
    $S_1$ is defined in (\ref{eq:Sk}):
    \begin{enumerate}[(a)]
    \item[(a)] \label{step:1a} Consider the observations in $t_L$. If $J >
      2$, form two superclasses, with one containing the class that
      minimizes the misclassification cost in $t_L$ and the other
      containing the rest.  Use Theorem~\ref{th:1} to obtain an
      ordering $a_1 \prec a_2 \prec \cdots$ of the values of $X_2$.
      Define $A_i = \{a_1, a_2, \ldots, a_i\}$, $t_{\mathit{LL}} = t_L \cap
      \{X_2 \in A_i\}$, and $t_{\mathit{LR}} = t_L \cap \{X_2 \notin A_i\}$
      for $i=1,2,\ldots.$
    \item[(b)] Repeat the preceding step on the observations in $t_R$ to
      obtain an ordering $b_1 \prec b_2 \prec \cdots$ of the values of
      $X_2$.  Define $B_j = \{b_1, b_2, \ldots, b_j\}$, $t_{\mathit{RL}} = t_R
      \cap \{X_2 \in B_j\}$, and $t_{\mathit{RR}} = t_R \cap \{X_2 \notin
      B_j\}$ for $j=1,2,\ldots.$
    \item[(c)] Let $\delta_1$ be the minimum value of~(\ref{eq:4nodes})
      over $\{A_i\}$, $\{B_j\}$, and $c \in S_1$. Let $c^*$ be the
      minimizing value of $c$.
    \end{enumerate}
  \item Find the following two splits of $t$ on $X_2$:
    \begin{enumerate}[(a)]
    \item[(a)] Order the values of $X_2$ in the set $\{X_1 \leq c^*\}$
      according to step~\ref{step:1a} above and use them to generate
      splits of the form $t_L = \{X_2 \in U_i\}$ and $t_R = \{X_2
      \notin U_i\}$, $i=1, 2, \ldots.$ For each $i$, find the best
      splits of $t_L$ and $t_R$ on $X_1$ into $t_{\mathit{LL}}$, $t_{\mathit{LR}}$ and
      $t_{\mathit{RL}}$, $t_{\mathit{RR}}$, respectively.  Let $\delta_2$ be the minimum
      value of~(\ref{eq:4nodes}), attained at $t_L = \{X_2 \in U\}$, say.
    \item[(b)] Repeat the preceding step on the observations in $\{X_1 >
      c^*\}$ to get the split $t_L = \{X_2 \in V\}$, say, with
      minimizing value $\delta_3$.
    \end{enumerate}
  \item If $\delta_1 \leq \min (\delta_2, \delta_3)$, split $t$ with
    $\{X_1 \leq c^*\}$. Otherwise, select $\{X_2 \in U\}$ if \mbox{$\delta_2
    \leq \delta_3$}, and $\{X_2 \in V\}$ if $\delta_2 > \delta_3$.
  \end{enumerate}
 \end{procedure}

\begin{procedure} \label{proc:splitcat2}
     Split selection between categorical variables $X_1$ and~$X_2$:
  \begin{enumerate}[3.]
  \item \label{step:cc:1} Given $U$, $V$ and $W$, let $t_L = \{ X_1
    \in U\}$, $t_R = \{ X_1 \notin U\}$, $t_{\mathit{LL}} = t_L \cap \{X_2 \in
    V\}$, $t_{\mathit{LR}} = t_L \cap \{X_2 \notin V\}$, $t_{\mathit{RL}} = t_R \cap
    \{X_2 \in W\}$, and $t_{\mathit{RR}} = t_R \cap \{X_2 \notin W\}$. Let
    $k_1$ be the number of distinct values of $X_1$:
    \begin{enumerate}[(a)]
    \item[(a)] If $J = 2$, search over all sets $U$, $V$ and $W$.
    \item[(b)] If $J > 2$ and $k_1 \leq 5$, search over all sets $U$ but
      restrict the sets $V$ and $W$ as follows.  Let $j_0$ be the
      class that minimizes the misclassification cost in $t$ and
      create two superclasses, with one containing $j_0$ and the other
      containing the rest. Use Theorem~\ref{th:1} on the two
      superclasses to induce an ordering of the $X_2$-values and then
      search for $V$ and $W$ using the ordered values.
    \item[(c)] If $J > 2$ and $k_1 > 5$, use the method in the preceding
      step to restrict the search on $U$, $V$ and $W$.
    \end{enumerate}
    Let $\delta_1$ be the smallest searched value
    of~(\ref{eq:4nodes}), and let it be attained at $U = U'$.
  \item Repeat step~\ref{step:cc:1} with the roles of $X_1$ and $X_2$
    reversed and let $\delta_2$ be the minimum value
    of~(\ref{eq:4nodes}), attained with $t_L = \{X_2 \in V'\}$.
  \item Split $t$ with $\{X_1 \in U'\}$ if $\delta_1 \leq \delta_2$;
    otherwise, split with $\{X_2 \in V'\}$.
  \end{enumerate}
 \end{procedure}
\end{appendix}

\section*{Acknowledgments}
The~author is grateful to editor M. Stein, an associate editor and a
referee for their helpful comments and suggestions. He also thanks
T.~Hothorn for assistance with the PARTY R package.

\printaddresses

\end{document}